\documentclass[12pt]{article}

\usepackage{sbc-template}

\usepackage{graphicx,url}
\usepackage{subfigure}

\usepackage[english]{babel}
\usepackage[utf8]{inputenc}  
\usepackage{multirow}
\usepackage{xcolor}
\usepackage{comment}

\usepackage{tikz}
\usepackage{pgfplots}
\usetikzlibrary{pgfplots.groupplots,patterns}
\pgfplotsset{
     compat=1.16, 
     compat/labels=pre 1.3 
}

\usepackage{float}

\makeatletter
\newcommand\resetstackedplots{
\makeatletter
\pgfplots@stacked@isfirstplottrue
\makeatother
}
\usepackage{multicol}
\usepackage{hyperref}
\usepackage[noend]{algorithm}  
\floatname{algorithm}{Algorithm}

\PassOptionsToPackage{dvipsnames}{xcolor} 
\usepackage{framed}
\usepackage[dvipsnames]{xcolor}

\definecolor{light-gray}{gray}{0.8}

\usepackage[noend]{algpseudocode}
\newcommand{\myand}{\textbf{and~}}
\newcommand{\myor}{\textbf{or~}}
\newcommand{\myfrom}{\textbf{from~}}
\newcommand{\myto}{\textbf{to~}}

\algblockdefx[Event]{Event}{EndEvent}%
[1]{\textbf{upon} #1}{}%
\algnewcommand\algorithmicfrom{\textbf{from}}
\algnewcommand\Recv[2]{\texttt{receive} #1 \algorithmicfrom\ #2}
\algnewcommand\FDEvent[2]{\textbf{notification} \texttt{#1}(#2)}

\makeatletter
\ifthenelse{\equal{\ALG@noend}{t}}%
    {\algtext*{EndEvent}}
    {}%
\algrenewcommand\ALG@beginalgorithmic{\small}
\makeatother

\usepackage{latexsym,amsmath,amsthm, amssymb,amsfonts}
\newtheorem{teo}{Theorem}
\newtheorem{lema}[teo]{Lemma}
\newenvironment{prooof}[1][Proof.]{\begin{trivlist}
\item[\hskip \labelsep {\bfseries #1}]}{\end{trivlist}}

\newcommand{\lista}[1]{\{#1\}}  
\newcommand{\ackTuple}[3]{\langle {#1}, {#2}, {#3}\rangle}
\newcommand{\ackElem}[3]{\{ \ackTuple{#1}{#2}{#3}\}}

\newcommand{\edge}[2]{({#1}, {#2})}

\usepackage{csquotes}

\title{\Large Distributed and Autonomic Minimum Spanning Trees}

\author{Luiz A. Rodrigues\inst{1}, Elias P. Duarte
Jr.\inst{2} and Luciana Arantes\inst{3}}

\address{
  Western Parana State University (UNIOESTE) --
  Cascavel -- PR -- Brazil
\nextinstitute 
  Federal University of Parana (UFPR) --
  Curitiba -- PR -- Brazil  
\nextinstitute 
  Sorbonne Université -- CNRS/LIP6 --
  Paris -- France  
\email{luiz.rodrigues@unioeste.br, elias@inf.ufpr.br, luciana.arantes@lip6.fr}
}

\begin{document}
\pagestyle{plain}

\maketitle

\begin{abstract}
The most common strategy for enabling a process in a distributed system to broadcast a message is one-to-all communication. However, this approach is not scalable, as it places a heavy load on the sender. This work presents an autonomic algorithm that enables the $n$ processes in a distributed system to build and maintain a spanning tree connecting themselves. In this context, processes are the vertices of the spanning tree. By definition, a spanning tree connects all processes without forming cycles. The proposed algorithm ensures that every vertex in the spanning tree has both an in-degree and the tree depth of at most $log_2 n$. When all processes are correct, the degree of each process is exactly $log_2 n$. A spanning tree is dynamically created from any source process and is transparently reconstructed as processes fail or recover. Up to $n-1$ processes can fail, and the correct processes remain connected through a scalable, functioning spanning tree. To build and maintain the tree, processes use the VCube virtual topology, which also serves as a failure detector. Two broadcast algorithms based on the autonomic spanning tree algorithm are presented: one for best-effort broadcast and one for reliable broadcast. Simulation results are provided, including comparisons with other alternatives.
\end{abstract}

\begin{displayquote}
\textbf{Note}: this preprint is an English translation and slightly extended version of the paper published in Portuguese at the 32nd Brazilian Symposium on Computer Networks and Distributed Systems (2014), reference \cite{rodrigues2014arvores}.
\end{displayquote}

\section{Introduction}

Spanning trees are used to solve various problems in distributed systems, such as mutual exclusion, clustering, message broadcasting, among many others \cite{ryan2024minimum, elkin:2006, england:2007, dahan:2009}. The main reason for being so important in this context is that a spanning tree represents a low-cost way to connect all processes of the system. 

Consider, for instance, message broadcasting. If a message is sent from the source to all destinations, as the number of processes grows, there is a high load on the source. The alternative is to allow processes relay messages to other processes, which relieves the source. However, depending on the topology the processes form, multiple copies of the message may be received by each process. A spanning tree avoids that, as it connects all processes without forming any cycles. For a message to reach all nodes it only needs to be transmitted along the $n-1$ edges of the tree \cite{gartner:2003}.

Another aspect that must be taken into account is the possibility of process failures, which is intrinsic to distributed systems. A fault-tolerant distributed application must continue its correct execution even after process faults have occurred. Thus, the spanning tree must be resilient to process failures.  Ideally, the failover procedure should be transparent. In this way, the system autonomically adapts itself to any fault situation \cite{kephart:2003}. Therefore, in addition to the algorithm employed to construct the tree, it is also paramount to have an algorithm for rebuilding or reconfiguring the tree as processes become faulty or recover.

In this work, we assume a fully-connected distributed system in which $n$ processes communicate through message passing. Processes may fail by crashing, and can also recover and rejoin the system. The processes form a virtual topology called the VCube \cite{duarte:1998,duarte2014vcube}. The topology is virtual in the sense that processes only communicate across the VCube edges. Nevertheless, given the underlying fully-connected topology any process can directly communicate with every other process, without employing intermediates. 

VCube is a failure detector \cite{chandra1996unreliable, reynal2005short,duarte2022distributed} that organizes processes into progressively larger hierarchical clusters. A VCube is a hypercube when the number of processes is a power of two and all processes are correct. The virtual topology presents several logarithmic properties that ensure its scalability. Those properties are guaranteed even as processes fail and recover 

We propose an autonomic spanning-tree algorithm constructed on a VCube in a distributed manner.  Starting from any process that is called the source and which becomes the tree root, each process communicates with the first correct process in its clusters.
Those processes in turn become roots of the subtrees in their clusters, and are connected to processes on their own clusters, and so on. When all processes are correct, the degree of a process is at most $log_2 n$, as well as the height of the spanning tree. A tree is reconstructed autonomously and transparently after failures and recoveries occur.

Two distributed broadcast applications were implemented using the proposed autonomic spanning trees: one for best-effort broadcast and the other for reliable broadcast. Simulation results confirm that the use of the virtual topology and the tree algorithm increases scalability in comparison with traditional one-to-all traditional alternatives. The message propagation latency is only slightly highly than that of a traditional all-communicates-with all strategy.

The rest of this work is organized as follows. The next section presents the system model and introduces the VCube topology. Section~\ref{sec:sta} presents the proposed spanning tree algorithm. Section~\ref{sec:apps} presents the tree-based broadcast algorithms. Simulation results are presented in Section~\ref{sec:simulation}. Section~\ref{sec:related} describes related work, in particular presenting a survey of the main distributed algorithms that have been proposed for the VCube. The conclusion follows in Section~\ref{sec:conclusions}.

\section{System Model \& The VCube Virtual Topology}\label{sec:sysmodel}

A distributed system consists of a finite set $\Pi$ of $n > 1$ independent
processes $\{p_0,...,p_{n-1}\}$ that collaborate to perform some
task. Each process is considered to be executed on a distinct node. Therefore, the terms {\em node} and {\em process} are used interchangeably. The
system under consideration is synchronous. Therefore, the temporal attributes related to process execution speed and message delay in communication channels
have known bounds.

Let $G=(V, E)$ be the complete and undirected graph representing $\Pi$, in which there are
$n=|V|$ vertices representing the processes and edge $\edge{i}{j}$ indicates that
process $i$ can communicate directly with process $j$ and vice versa. The graph is complete, and represents a fully connected system, in which any process can directly communicate with any other process. Thus $\exists \edge{i}{j}, \forall i, j \in V, i \neq j$. A (\textit{spanning tree}) of $G$ is a connected and acyclic subgraph $T=(V, E')$ in which $E'\subseteq E$, that is, $T$, contains all the vertices of $G$ and $|E'| = |V|-1$. If edges have associated weights, a \textit{minimum spanning tree} is one whose sum of the edge weights is minimal. If each edge has a different weight, there is a single minimal tree. If all edges have the same weight, all trees in the graph are minimal \cite{gallager:1983}. In this work, all links have the same weight which is equal to 1 and omitted in the rest of the paper.

Processes communicate by sending and receiving messages. The processes are organized in a virtual hypercube. A $d$-dimensional hypercube has $n=2^d$ processes. The identifier $i$ of a process satisfies $0 \leq i \leq n-1$ and consists of a $d$-bit binary address $i_{d-1}, i_{d-2}, ..., i_0$. Two processes are connected if their addresses differ by exactly one bit. The transmission and receipt of messages are atomic operations, but {\it multicast} and {\it broadcast} are not supported as native operations. The links are reliable and never fail. Therefore, no messages are lost, corrupted, or duplicated during transmission. There is no network partitioning.

Processes can fail by crashing, and such failures are permanent. A process is considered to be ``correct'' or ``fault-free'' if it does not fail during the entire system execution. Otherwise, the process is considered to be ``faulty''. After it has crashes, a process does not perform any actions and does not respond to external stimuli; that is, it stops executing completely, and does not send or receive messages.

Faulty processes are detected by a perfect failure detector; that is, no faulty process $q$ is suspected unless it is actually faulty, and if $q$ is faulty, every correct process $p$ will be notified by the detector module about the failure of $p$ within a finite time \cite{freiling:2011}.

Using a virtual topology to connect processes in a distributed system
facilitates application development by abstracting the physical structure and facilitating system reconfiguration when necessary.
The topology employed in this work, called VCube, organizes the
processes into a virtual hypercube when there are no failures and the number of processes is a power of 2. However, if a process fails, the VCube reconfigures itself to adapt and maintain the topology connected. Even in the presence of failures, VCube maintains the following logarithmic properties: in a $d$-dimensional VCube with $n=2^d$
vertices, the total number of edges is at most $n \log_2 n$ despite of which processes are 
faulty or correct, and the maximum distance between two vertices is $\log_2 n$.

\begin{figure}[htb]
  \centering
  \includegraphics[scale=0.9]{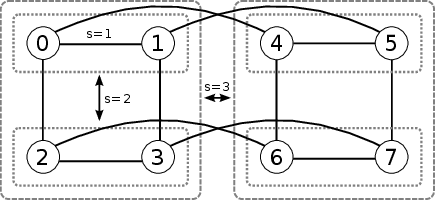}
  \caption{The clusters of a three-dimensional VCube.}
  \label{fig:hiadsd}
\end{figure}

To build the VCube processes are organized into progressively larger clusters \cite{duarte:1998}, as illustrated in Figure~\ref{fig:hiadsd}. Each process $i$ has $s=1, .., \log_2 n$ clusters each with $2^{s-1}$ processes. Clusters in which $s=1$ have one process. Clusters of in which $s=2$ have 2 processes, and so on.

The processes belonging to each cluster can be computed with function $C_{i, s}= \lista{i \oplus 2^{s-1}, c_{i \oplus
2^{s-1}, 1}, ..., c_{i \oplus 2^{s-1}, s-1}}$, in which $\oplus$ represents the binary operation of {\it exclusive or} ({\it xor}).

Let $i$ and $j$ be two processes of the system. Next, we define three functions on the VCube: $cluster_i(j)$, $FF\_neighbor_i(s)=j$, and $neighborhood_i(h)$, where $h$ corresponds to the height of spanning tree as will be detailed below.

Function $cluster_i(j) = s$ determines
which cluster $s$ of process $i$ process $j$ belongs to. For example, in Figure~\ref{fig:hiadsd} $cluster_0(1)=1$,
$cluster_0(2)=cluster_0(3)=2$, $cluster_0(5)=3$. Note that for every pair $i,j$,
$cluster_i(j) = cluster_j(i)$.

Another function defined on the VCube is $FF\_neighbor_i(s)=j$. This function computes the first correct process $j$ in cluster $C_{i,s}$. If all processes in the cluster
are faulty, the function returns $\bot$. In the VCube in Figure~\ref{fig:hiadsd},
for example, in a fault-free scenario $FF\_neighbor_4(1)=5$,
$FF\_neighbor_4(2)=6$, $FF\_neighbor_4(3)=0$. On the other hand, if $p_4$
has detected $p_6$ faulty, $FF\_neighbor_4(2)=7$. If $p_6$ and $p_7$
are faulty, $FF\_neighbor_4(2)=\bot$. It is important to emphasize that this function is
dependent on which processes have been detected as faulty.
Due to the latency required for detection, within the same timeframe, it is possible
for two processes to have different views on which processes are correct or faulty.

Finally, function $neighborhood_i(h) = \{j~|~j = FF\_neighbor_i(s), j \neq
\bot, 1 \leq s \leq h\}$ is defined.
This function generates a set containing all correct processes
virtually adjacent to process $i$ according to $FF\_neighbor_i(s)$, for
$s=1, .., h$. Parameter $h$ varies from 1 to $\log_2 n$. If $h=\log_2 n$ the resulting set contains all correct neighbors of process $i$. For any other value of $h < \log_2 n$, the function returns only a subset of the neighbors contained in the clusters $s=1, .., h$. Like $FF\_neighbor_i$, this function depends on the local knowledge that
process $i$ has about the state of the other processes. As an example, for the
VCube in Figure~\ref{fig:hiadsd} and in a fault-free scenario,
$neighborhood_0(1)=\{1\}$, $neighborhood_0(2)=\{1,2\}$, and
$neighborhood_0(3)=\{1,2,4\}$. If process $p_4$ is detected as faulty
by process $p_0$, then $neighborhood_0(3)=\{1,2,5\}$.
If $p_1$ is faulty, $neighborhood_0(1)=\emptyset$.

Thus, the system topology in VCube is formed by the connection of each process
$i \in V$ with all its neighbors as determined by function $neighborhood_i(\log_2 n)$.
Similarly, the neighbors of a process $i$ restricted to the cluster $s$ to which
$i$ belongs in relation to another process $j$ are determined by
$neighborhood_i(cluster_i(j)-1)$.

\section{The Spanning Tree Algorithm}\label{sec:sta}

This section presents the proposed autonomic and distributed spanning tree algorithm based on the VCube topology. The spanning tree is considered to be \textit{autonomic} because it dynamically rebuilds itself as processes crash and recover. Moreover, in such cases the tree is regenerated locally, by the affected neighbor processes.

Consider that the distributed system is represented as a graph, of which the $n$ processes are vertices. The spanning tree connects all those processes using the minimum number of edges: $n-1$. Algorithm~\ref{alg:tree.distr} shows how a message called TREE is propagated from the source process (which represents the root of the spanning tree) to all other correct processes.

\begin{algorithm}[htb]
\caption{Propagation of Message TREE along the Spanning Tree}
\label{alg:tree.distr}
  \begin{algorithmic}[1]
   \State $correct_i \gets \{0, .., n-1\}$ // list of correct processes
 
  \vspace{6pt}
  \Procedure{StartTree}{ }
     \State // executed by the source, which is the tree root
     \State // the source sends message TREE to all neighbors
      \ForAll{$k \in neighborhood_i(\log_2 n)$}\label{alg:tree.distr.starttree}
           \State \Call{Send}{$\langle TREE\rangle$} to $p_k$
     \EndFor\label{alg:tree.distr.starttree1}    
  \EndProcedure
  
  \vspace{6pt}  
  \Event{\Recv{$\langle TREE\rangle$}} \myfrom $j$
    \State // a process that is not the source receives message TREE
    \If {$j \in correct_i$} 
       \State // retransmits to the neighbors in internal clusters
       \For{$k \in neighborhood_i(cluster_i(s)-1)$}
\label{alg:tree.distr.subtree}
          \State \Call{send}{$\langle TREE\rangle$} to $p_k$  
       \EndFor
    \EndIf
  \EndEvent  
  
  \vspace{6pt}
  \Event{\FDEvent{crash}{process $j$}}\label{alg:tree.distr.crash} 
    \State // process $i$ is notified that process $j$ has been detected as crashed
    \State $correct_i \gets correct_i \smallsetminus \{j\}$
    \If {$k = FF\_neighbor_i(cluster_i(j)), k \neq
\bot$}\label{alg:tree.distr.recover}
         \State \Call{send}{$\langle TREE\rangle$} to $p_k$
    \EndIf 
   \EndEvent 
 \end{algorithmic}
\end{algorithm}

Initially, consider a fault-free execution in which all processes are correct. The \textsc{StartTree} procedure is executed by the source process, which is the tree root. Note that any process can be the source. A TREE message is sent to
the root's $\log_2 n$ correct neighbors, one in each cluster $s=1, .., \log_2 n$
(lines~\ref{alg:tree.distr.starttree}-\ref{alg:tree.distr.starttree1}). 

Upon receiving message TREE from process $j$, process $i$ forwards the message to its internal clusters, with respect to $j$ and $cluster_i(j)$. Thus $i$ forwards message TREE to the first correct process in clusters $s'=1, .., cluster_i(j)-1$
(line~\ref{alg:tree.distr.subtree}).

\begin{figure}[htb]
  \centering
  \subfigure[fault-free]
     {\label{fig:st3}%
      \includegraphics[scale=0.7]{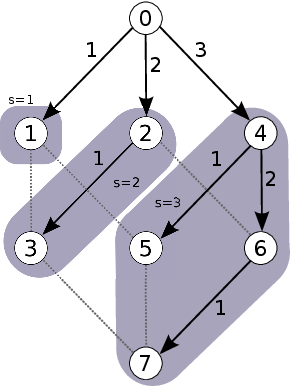}} 
  \hfil
  \subfigure[process $p_4$ is crashed]
     {\label{fig:st3.falha}%
      \includegraphics[scale=0.7]{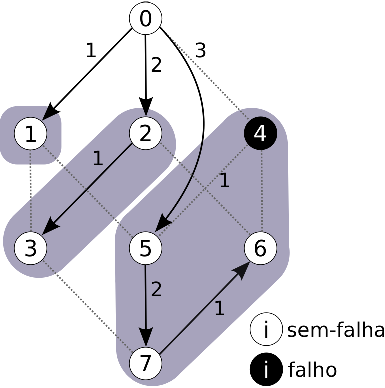}}
  \caption{An example spanning tree built on a 3-dimensional VCube.}
  \label{fig:adsd.diagnostico}
\end{figure}

Figure~\ref{fig:st3} shows as an example a 3-dimensional VCube in which all processes are correct. 
Process $p_0$ is the source and sends message TREE to its neighbors, the first correct process in each of its $log n$ clusters, computed as follows: $neighborhood_0(3)=\{FF\_neighbor_0(1), FF\_neighbor_0(2), FF\_neighbor_0(3)\} = \{1,2,4\}$. Process $p_1$ receives the message but does not retransmit it, since $cluster_1(0)=1$ and
$neighborhood_1(0)=\bot$. Process $p_2$ receives the message and forwards it to its neighbor $p_3$ in cluster $s=1$, since $neighborhood_2(1) = \{FF\_neighbor_2(1)\} = \{3\}$. Similar to $p_1$, When $p_3$ receives the message, it
computes $cluster_3(2)=1$ and does not retransmit it. In the case of $p_4$, the message
is received and retransmitted to neighbors $5 \in c_{4, 1}$ and $6 \in c_{4,2}$.
Finally, since $FF\_neighbor_6(1)=7$, process $p_6$ sends the message
to process $p_7$.

Failure cases can be divided into two scenarios. First, consider that a process $j \in c_{i,s}$ has crashed and process $i$ has already been
informed of this fault by the detector, that is, $j \notin correct_i$. In this case, process $i$ sends the message to its correct neighbor
$k=FF\_neighbor_i(s)$ and the message is propagated correctly by $k$ to all
its internal clusters (of $k$). In a second scenario, consider the same crashed process $j$, but process $i$ has not yet been informed by the detector, that is, $j
\in correct_i$. In this case, if $FF\_neighbor_i(s)=j$, process $i$ sends the $TREE$  message to $j$, and thus the message is not received. Propagation stops prematurely and the
internal subtree of cluster $s$ does not get the message. However, as soon as the detector module informs $i$ about the failure, a new $FF\_neighbor_i(s)=k$ is selected and the message is retransmitted to $k$, thus rebuilding the subtree and allowing the message to reach all processes (line~\ref{alg:tree.distr.recover}).

Figure~\ref{fig:st3.falha} illustrates a failure scenario. Let process
$p_0$ be the root and $p_4$ be the crashed process. Since $p_0$ is aware of 
the failure of process $p_4$, instead of sending message TREE to $p_4$, 
$p_0$ sends the message to $p_5$, which is the first correct process in $c_{0,3}$. 
Process $p_5$, in turn, forwards the message to $p_7 \in c_{5,2}$. Finally, $p_7$ retransmits the message to $p_6 \in c_{7,1}$, completing the tree. 
If when $p_0$ starts the broadcast and $p_4 \in correct_i$, $p_0$ sends the message to $p_4$ and, when the detector informs it about the failure of $p_4$, the message will be
retransmitted to $p_5$. From this point on, propagation is analogous to the previous case.

Lemma~\ref{lemma:reception_correct} proves that all correct processes receive the a message that is propagated across the tree built with Algorithm~\ref{alg:tree.distr}. In the next section, two broadcast algorithms based on the spanning tree are proposed, and this lemma is used in the proof of their correctness.

\begin{figure}[htb]
  \centering \includegraphics[scale=1.0]{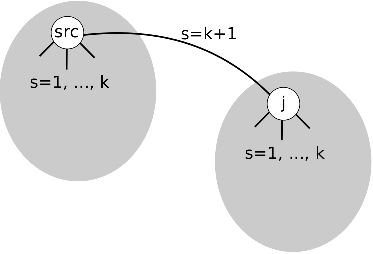}
  \caption{Dissemination of TREE messages in a system with $n=2^{k+1}$ processes which consists of two subsystems with $n=2^k$ processes.}
\label{fig:proof.neighbor}
\end{figure}

\vspace{6pt}
\begin{lema}\label{lemma:reception_correct}
Let $m$ be a message propagated by a correct source process $src$ through the spanning tree. Every correct process receives $m$.  
\end{lema} 
\begin{prooof}

The proof of this lemma is by induction. Consider as the basis of induction a system with $n=2$ processes: $p_0$ is the $src$ process that initiates the transmission of message $m$ across the tree. According to the $c(i,s)$ function, $p_1 \in c_{0,1}$. If $p_1$ is correct, $FF\_neighbor_0(1)=1$ and $p_0$ sends $m$ to $p_1$ (line \ref{alg:tree.distr.starttree}). Therefore, $p_1$ receives $m$ and
the lemma holds.

As the induction hypothesis, consider that the lemma holds for a system with
$n=2^k$ processes.

The induction step demonstrates that the lemma holds for a system with
$n=2^{k+1}$ processes. According to VCube's hierarchical topology, a system with $n=2^{k+1}$ processes consists of two subsystems with $n=2^k$ processes, as illustrated in Figure~\ref{fig:proof.neighbor}. The figure shows that $src$ and $j$ are the roots of those subsystems. Process $src$ executes the algorithm and sends $m$
to each process returned by $FF\_neighbor_{src}(s)$, $s=1,..,k$ (line
\ref{alg:tree.distr.starttree}). Since $j = FF\_neighbor_{src}(k)$ is a correct
process, $j$ correctly receives $m$. If $j$ is detected as crashed, a copy
of the message is retransmitted to the next correct process in the same cluster as
$j$ (line~\ref{alg:tree.distr.recover}). Thus, message $m$ is transmitted
in both subsystems (of $n=2^k$ processes) and, by hypothesis, every correct process receives $m$ in each subsystem. Since every process in $\Pi$ belongs to one of those systems, every correct process in $\Pi$ receives $m$.

\end{prooof}

\section{Broadcast Algorithms Based on the Spanning Tree}\label{sec:apps}

Message broadcasting \cite{hadzilacos:1993} is a fundamental building block used to implement several distributed algorithms. Examples include content delivery networks, publish/subscribe systems, distributed replication, and group communication. A source process broadcasts a message to all other processes in the system. However, if the source fails before completing the message transmission, some destination processes may not receive the message. If the required type of broadcast is \textit{best-effort}, nothing further is needed; this ensures that only if the source is correct, all correct processes deliver the message. In contrast, if it is necessary to guarantee that all correct processes deliver the message even if the source fails, a different type of broadcast is required: \textit{reliable broadcast}.

Fault-tolerant broadcast algorithms are typically implemented assuming underlying reliable point-to-point communication channels between processes. Processes communicate by sending and receiving messages executing the {\sc send} and {\sc receive} primitives. Processes invoke {\sc broadcast}($m$) and {\sc deliver}($m$) to broadcast and deliver messages to/from other application processes. A broadcast algorithm may rely on a failure detector to be notified if some process fails.

Next, two broadcast algorithms based on the VCube are presented. The first is a best-effort broadcast algorithm, while the second implements reliable broadcast. Both algorithms disseminate messages through the spanning tree described in Section~\ref{sec:sta}.

\subsection{Best-Effort Broadcast}\label{sec:besteffort}

Best-effort broadcast guarantees that all correct processes deliver the same set of messages as long as the sender (source) is correct. A best-effort broadcast algorithm must guarantee three properties: validity, no duplication, and
no creation of messages \cite{guerraoui:2006}. According to the validity property,
if a process $p_i$ sends a message $m$ to a process $p_j$ and neither
of them fails, then $p_j$ eventually delivers $m$. No duplication guarantees that
no message is delivered more than once, and no creation guarantees that no
message is delivered unless it has been previously broadcast by some correct process.

Algorithm~\ref{alg:brcast} is a best-effort broadcast solution on the VCube spanning tree. Two message types are used: $\langle TREE, m\rangle$ for application messages and $\langle ACK, m\rangle$ which are acknowledgment messages (ACKs) used to confirm the receipt of application messages. The VCube failure detector notifies correct processes whenever some process is detected to have crashed. The algorithm works correctly even if up to $n-1$ processes crash. A preliminary version of this algorithm was published by Rodrigues \cite{rodrigues:2013} applied to distributed mutual exclusion.

Processes executing the best-effort broadcast algorithm maintain the following local variables:

\begin{itemize}
    \item $correct_i$: set of processes that process $i$ considers to be correct;
    
    \item $last_i[n]$: the last message received from each source process.  Every message has a unique identifier, consisting of the source process id plus a timestamp. The timestamp is implemented by the source as local counter of messages. Function $ts(m)$ returns the timestamp of message $m$. Note that any process will only send a new message after it was successfully completed the broadcast of the previous message;
    
    \item $ack\_set_i$: the set of ACKs that process $i$ is waiting for (called pending ACKs). For each message $\langle TREE, m\rangle$ received by process $i$ from process $j$ and retransmitted to process $k$, an element $\ackTuple{j}{k}{m}$ is added to this set.
\end{itemize}

The symbol $\bot$ represents a null element. The asterisk is used as a wildcard
to select ACKs from the set $ack\_set$. Thus, for example: an element such as 
$\ackTuple{j}{*}{m}$ represents all pending ACKs for message $m$ received from 
process $j$ and retransmitted to any other process.

\begin{algorithm}[H]
\caption{The Hierarchical Best-Effort Broadcast Algorithm executed by process $i$}\label{alg:brcast}
\scriptsize
  \begin{algorithmic}[1]
    \State $last_i[n] \gets \{\bot, .., \bot\}$
    \State $ack\_set_i = \emptyset$
    \State $correct_i = \{0, .., n-1\}$
  
  \vspace{6pt}
   \Procedure{Broadcast}{message $m$}
     \State \textbf{wait until} $ack\_set_i~\cap~\ackElem{\bot}{*}{last_i[i]} = \emptyset$\label{alg:brcast.end}
     \State $last_i[i] = m$
     \State \Call{Deliver}{$m$}\label{alg:brcast.deliver1}
    \ForAll{$j \in neighborhood_i(\log_2 n)$}\label{alg:brcast.2allneighbor} // sends the message to all neighbors
       \State $ack\_set_i \gets  ack\_set_i~\cup \ackElem{\bot}{j}{m}$
       \State \Call{Send}{$\langle TREE, m\rangle$} \myto $p_j$\label{alg:brcast.2allneighbor2}
    \EndFor
   \EndProcedure
  \vspace{6pt}

  \Procedure{CheckAcks}{process $j$, message $m$}
     \If{$ack\_set_i~\cap \ackElem{j}{*}{m} = \emptyset$}
        \If{$\{source(m), j\} \subseteq correct_i$}
           \State \Call{Send}{$\langle ACK, m\rangle$} \myto $p_j$\label{alg:brcast.sendack}
        \EndIf
    \EndIf
  \EndProcedure

  \vspace{6pt}
  \Event{\Recv{$\langle TREE, m\rangle$}} \myfrom $p_j$\label{alg:brcast.recv.tree}
    \If{$\{source(m), j\} \nsubseteq correct_i$}
       \State \Return{}
    \EndIf   
    \State // checks if $m$ is a new message
    \If{$last_i[source(m)]=\bot$ \myor $ts(m) = ts(last_i[source(m)])+1$ 
\label{alg:brcast.noduplication}} 
       \State $last_i[source(m)] \gets m$
       \State \Call{Deliver}{$m$}\label{alg:brcast.deliver2}
    \EndIf
    \State // retransmits $m$ to neighbors in internal clusters
    \ForAll{$k \in neighborhood_i(cluster_i(j)-1)$} \label{alg:brcast.fw} 
        \If {$\ackTuple{j}{k}{m} \notin ack\_set_i $}
           \State $ack\_set_i \gets  ack\_set_i~\cup \ackElem{j}{k}{m}$
           \State \Call{Send}{$\langle TREE, m\rangle$} \myto $p_k$
        \EndIf
    \EndFor
    \State \Call{CheckAcks}{$j$, $m$}\label{alg:brcast.check.tree}
  \EndEvent
  
  \vspace{6pt}
  \Event{\Recv{$\langle ACK, m\rangle$}} \myfrom $p_j$\label{alg:brcast.rcv.ack}
    \State $k \gets x:\ackTuple{x}{j}{m} \in ack\_set_i$
    \State $ack\_set_i \gets ack\_set_i~\smallsetminus~\ackElem{k}{j}{m}$
    \If {$k \neq \bot$} 
        \State \Call{CheckAcks}{$k$, $m$}\label{alg:brcast.check.ack}
    \EndIf
  \EndEvent
  
  \vspace{6pt}
  \Event{\FDEvent{crash}{process $j$}}\label{alg:brcast.crash} // $j$ is detected as crashed
    \State $correct_i \gets correct_i \smallsetminus \{j\}$
    \ForAll{$p=x, q=y, m=z:\ackTuple{x}{y}{z} \in ack\_set_i$}
      \If {$\{source(m), p\} \nsubseteq correct_i$} 
         \State // remove pending ACKs for $\ackTuple{j}{*}{*}$ and $\ackTuple{*}{*}{m}: source(m)=j$
         \State $ack\_set_i \gets ack\_set_i \smallsetminus \{\langle p, q, m\rangle\}$
      \ElsIf{$q = j$} // retransmits to new neighbor $k$, if exists
         \State $k \gets FF\_neighbor_i(cluster_i(j))$\label{alg:brcast.nextk}
         \If {$k \neq \bot$ \myand  $\ackTuple{p}{k}{m} \notin ack\_set_i$}
            \State $ack\_set_i \gets  ack\_set_i~\cup \ackElem{p}{k}{m}$
            \State \Call{Send}{$\langle TREE, m\rangle$} \myto $p_k$
         \EndIf
         \State $ack\_set_i \gets ack\_set_i \smallsetminus \{\langle p, j, m\rangle\}$
         \State \Call{CheckAcks}{$p$, $m$}
      \EndIf
    \EndFor
  \EndEvent  
  \end{algorithmic}
\end{algorithm}

\subsubsection{Algorithm Description}

A process $i$ running Algorithm~\ref{alg:brcast} invokes the {\sc Broadcast} primitive to broadcast a message $m$. A new message is broadcast only after the previous one has completed (line~\ref{alg:brcast.end} ), that is, when there are no more pending ACKs for the message $last_i[i]$. In lines~\ref{alg:brcast.2allneighbor}-\ref{alg:brcast.2allneighbor2} the new message is sent to all neighbors $j$ considered to be correct. For each message sent to $j$, a tuple $\ackTuple{\bot}{j}{m}$ is added to the list of pending ACKs.

When process $i$ receives a TREE message from process $j$
(line~\ref{alg:brcast.recv.tree}), it first checks whether both the
message source and $j$ are correct. If one of those processes has crashed, the reception is aborted. If $j$ has crashed, the process that had sent $m$ to $j$ will make resend that message after it detects that $j$ has crashed. Then process $i$ will receive the message through after the tree is rebuilt. On the other hand, if the source crashes, it is no longer necessary to continue the broadcast. If both the source and $j$ are correct, process $i$ checks whether the message is new by making a comparison of its {\it timestamp} with that of the last message stored in $last_i[j]$ and the received message $m$. If the message is a new one, then $last_i[j]$ is updated, and the 
the message is delivered to the application. Next, $m$ is forwarded to the
neighbors in each cluster of $i$. If there is no correct neighbor
or if $i$ is a leaf in the tree ($cluster_i(j)=1$), there are no pending ACKs ({\sc CheckAcks}) then $i$ immediately sends an ACK to $j$.

If a message $\langle ACK, m\rangle$ is received (line~\ref{alg:brcast.rcv.ack}), set $ack\_set_i$ is updated, and if there are no more pending ACKs for message $m$, process $i$ sends a $\langle ACK, m\rangle$ to the process $k$ from which $i$ previously received the TREE message ({\sc CheckAcks}). However, if $k = \bot$, the ACK has reached the source process and no longer needs to be propagated.

The detection of a crashed process $j$ is handled by the {\sc Crash} event. Three
actions are performed: (1) updating the list of correct processes; (2) removing pending ACKs for messages received from $j$ or messages that contain process $j$ as the destination; (3) messages that had been previously transmitted to $j$ must now be retransmistted to another correct process $k$ in the same cluster as $j$, if there is one. This retransmission triggers the propagation the message along a new subtree.

In Theorem~\ref{theo:besteffort}, the correctness of the proposed best-effort broadcast algorithm is proven. 

\vspace{3pt}
 \begin{teo}\label{theo:besteffort}
Algorithm~\ref{alg:brcast} ensures that if the source process remains correct, every correct process will eventually deliver the broadcast message.
 \end{teo}
 
 \begin{prooof}
The no duplication and no creation properties are derived from the underlying communication channels, which are perfect (reliable). Furthermore, every message has a unique identifier (timestamp), and even if some process receives that message multiple times, it is only delivered once (line~\ref{alg:brcast.noduplication}).

Reliable delivery is guaranteed as follows. After the source broadcasts message $m$, it waits for acknowledgment messages (ACKs) from all correct neighbors. Lemma~\ref{lemma:reception_correct} proved that if the sender is correct, every correct process receives $m$, even if an intermediate process $j$ crashes during retransmission. In this case, the message is retransmitted to the next correct neighbor $k$ in the same cluster as process $j$ (line~\ref{alg:brcast.nextk}).
\end{prooof}

\begin{teo}

The total number of messages generated in a fault-free execution of Algorithm~\ref{alg:brcast} is $2*(n-1)$, including ACKs. If a process $i$ that
received a message from a process $j$ crashes before sending back the ACK, the
total number of extra messages depends on the number of processes in the cluster and their states. Recall that a cluster consists of $2^{s-1}$ processes and the largest cluster has $n/2$ processes. 

\end{teo}

\begin{prooof}

In a fault-free execution of Algorithm~\ref{alg:brcast}, for each TREE message sent, an ACK message is returned. If $n-1$ is the number of edges in the tree that has $n$ processes as vertices, then the total number of messages is twice the total number of edges.

If a process $j$ is detected to have crashed by a process $i$ after it has sent a
TREE message was $j$, a new message will be sent to the next neighbor $k=FF\_neighbor_i(cluster_i(j))$. Let $s=cluster_i(j)$. In the best case, $k=\bot$ and no extra messages are sent (this is the case if $j \in c_{i,1}$ or there are no more correct processes in cluster $s$). However, if $k \neq \bot$, the number of extra messages depends on the number of processes detected as crashed/correct in cluster $s$. Let $n'=|c_{i,s}|$ denote the total number of processes in cluster $c_{i,s}$. In the worst case, all processes in the cluster are correct except $j$, and $j$ has sent the message to all neighbors before it crashed. Thus, in this case the total number of extra messages is $1+2*(n'-2)$. An extra TREE message is sent for $k$ and $(n'-2)$ TREE messages
+ $(n'-2)$ ACKs are sent in the subtree. In general, if there are $f$ faulty processes
in $c_{i,s}$ including $j$, the number of extra messages retransmitted is
$1+2*(n'-1-f)$.

\end{prooof}

\subsection{Reliable Broadcast}

A reliable broadcast algorithm ensures that the same set of messages is delivered to all correct processes, even if the source crashes as broadcast is being executed. The reliable broadcast algorithm proposed in this work uses the VCube spanning trees described in Section~\ref{sec:sta} to propagate messages hierarchically, and it works even if up to $n-1$ processes crash. Algorithm~\ref{alg:rb} is similar to the best-effort broadcast Algorithm \ref{alg:brcast}, but must handle the crash of the source process differently. Algorithm~\ref{alg:rb} only shows the difference. The rest of the pseudo-code is identical and has been omitted.

A process $i$ that is the broadcast source invokes primitive {\sc
Broadcast}($m$). If $i$ never crashes (in this case $i$ is the message source, i.e. $source(m)=i$) the algorithm is identical to the best-effort algorithm. After delivering the message locally, $i$ sends the message to the first correct neighbor in each of its clusters. If $i$ is not the source of the message, it only retransmits the message to its neighbors.

When a process $i$ receives the message $\langle TREE, m\rangle$ from a process
$j$, it first checks whether $j \notin correct_i$. If $j$ has crashed, the
message is discarded. Note that this check differs from that of the
best-effort algorithm as the reliable broadcast algorithm does not discard messages received from a crashed $source(m)$. The second difference is in
lines~\ref{alg:rb.sourcefaulty.begin}-\ref{alg:rb.sourcefaulty.end}. If $i$
receives a new message from a source process which has crashed, it initiates a
new broadcast with the received message to ensure that the other
processes receive $m$ correctly. In this case, in line~\ref{alg:rb.sourcem},
$source(m) \neq i$ and the message is rebroadcast to the other processes through the
spanning tree of $j$.

Crash notifications are handled similarly to best-effort broadcast, except that is necessary to rebroadcast the last message received from the process notified to have crashed (line~\ref{alg:rb.aftercrash}). The
retransmissions in line~\ref{alg:rb.crashsource} ensure that all correct processes will receive and deliver the last message broadcast by the crashed source process $j$. That is guaranteed even if a single correct process received the message before $j$ crashed.

\begin{algorithm}[H]
\caption{Hierarchical Reliable Broadcast executed by process $i$}\label{alg:rb}
\scriptsize
  \begin{algorithmic}[1]
  \vspace{6pt}
   \Procedure{Broadcast}{message $m$}
     \If {$source(m) = i$}\label{alg:rb.sourcem}
        \State \textbf{wait until} $ack\_set_i~\cap~\ackElem{\bot}{*}{last_i[i]} = \emptyset$\label{alg:rb.end}
        \State $last_i[i] = m$
        \State \Call{Deliver}{$m$}\label{alg:rb.deliver1}
     \EndIf
     \State // sends the message to all neighbors
     \State ...
   \EndProcedure

   \vspace{6pt}
    \Event{\Recv{$\langle TREE, m\rangle$}} \myfrom $p_j$\label{alg:rb.recv.tree}
      \If{$j \notin correct_i$}
         \State \Return{}
      \EndIf
     \State // checks if $m$ is a new message
     \If{$last_i[source(m)]=\bot$ \myor\\ 
        \hspace{20pt}$ts(m)
=ts(last_i[source(m)])+1$}\label{alg:rb.noduplication}
       \State $last_i[source(m)] \gets m$
       \State \Call{Deliver}{$m$}\label{alg:rb.deliver2}
        \If {$source(m) \notin correct_i$}\label{alg:rb.sourcefaulty.begin}
           \State \Call{Broadcast}{$m$}\label{alg:rb.crashsource}
           \State \Return{}
        \EndIf\label{alg:rb.sourcefaulty.end}
     \EndIf
     \State ...
   \EndEvent
  
  \vspace{6pt}
 \Event{\FDEvent{crash}{process $j$}}\label{alg:rb.crash} // $j$ is detected as crashed
    \State $correct_i \gets correct_i \smallsetminus \{j\}$

    \ForAll{$p=x, q=y, m=z:\ackTuple{x}{y}{z} \in ack\_set_i$}
      \If {$p \notin correct_i$} 
         \State // remove pending ACKs for $\ackTuple{j}{*}{*}$
         \State $ack\_set_i \gets ack\_set_i \smallsetminus \{\langle p, q, m\rangle\}$
      \ElsIf{$q = j$} // retransmits $m$ to new neighbor $k$, if exists
         \State $k \gets FF\_neighbor_i(cluster_i(j))$\label{alg:rb.nextk}
         \If {$k \neq \bot$ \myand  $\ackTuple{p}{k}{m} \notin ack\_set_i$}
            \State $ack\_set_i \gets  ack\_set_i~\cup \ackElem{p}{k}{m}$
            \State \Call{Send}{$\langle TREE, m\rangle$} \myto $p_k$
         \EndIf
         \State $ack\_set_i \gets ack\_set_i \smallsetminus \{\langle p, j, m\rangle\}$
         \State \Call{CheckAcks}{$p$, $m$}
      \EndIf
    \EndFor
    \If{$last_i[j] \neq \bot$}
       \State \Call{Broadcast}{$last_i[j]$}\label{alg:rb.aftercrash}
    \EndIf    
  \EndEvent  
  \end{algorithmic}
\end{algorithm}

\section{Simulation}\label{sec:simulation}

The proposed broadcast algorithms were implemented using Neko \cite{neko:2002}, a Java framework for simulating distributed algorithms. The proposed hierarchical best-effort broadcast algorithm is called ATREE-B and the reliable broadcast algorithm ATREE-R. The algorithm was compared with two other approaches:  the first is based on the classic all-to-all approach, and is called ALL-B (best-effort broadcast) and ALL-R (reliable broadcast). The second approach that was compared is also tree-based, but non-autonomic: NATREE-B (best-effort broadcast) and NATREE-R (reliable broadcast). We refer to the three different approaches as ATREE, ALL, and NATREE. All algorithms employ the failure detection service provided by VCube to learn about process crashes.

The ALL strategy also employs a $neighborhood_i$ function, but in this case it includes \textit{all} correct neighbors of process $i$. Whenever process $i$ disseminates a message, it sends the message to all those neighbors. The NATREE strategy builds a tree using flooding. The tree is created using the TREE broadcast message itself. Each process that receives the message for the first time joins the tree and retransmits the message to the other processes in the system, except the one from which it received the message. If a
process that receives the message is already in the tree, it just sends back a
NACK message and does not forward the message. Once the tree is created,
further messages are sent only along the edges of the tree. In case of failure,
a new tree is created from the source process using the same
flooding procedure.

\subsection{Simulation Parameters}

When a process needs to send a message to more than one recipient, it must execute the {\sc Send} primitive sequentially. Thus, for each message, $t_s$ time units are used to send the message and $t_r$ time units to receive it, in addition to the transmission delay $t_t$.
These intervals are computed for each copy of the message sent.

To evaluate the performance of the broadcast algorithms, two metrics are used: (1) \textit{throughput}, given by the total number of broadcasts completed during a time interval; (2) the \textit{latency} to deliver the broadcast message to all correct processes.

The proposed algorithms were evaluated in different scenarios, varying both the total number of processes and the number of crashed processes. The communication parameters were set to $t_s=t_r=0.1$ and $t_t=0.8$. The failure detector employs a testing interval set to $5.0$ time units. A process is considered to have crashed if it does not respond to the test after $4*(t_s+t_r+t_t)$ time units (timeout interval).

\subsection{Simulation Results}

The experiments were conducted in two stages. Initially, fault-free scenarios were used with systems with 8 to 1,024 processes. Furthermore, the latency
and the number of messages sent by each application were computed individually in a system with 512 processes. Subsequently, failure scenarios were randomly generated for a system with 512 processes containing from 1\% to 8\% of crashed processes.

\noindent{\bf Experiments in fault-free scenarios.} In fault-free scenarios,
since there are no retransmissions, the proposed VCube-based best-effort and reliable broadcast algorithms present very similar performance. 
Figure~\ref{fig:graph.ff} shows the latency and throughput considering
that a single message is sent by process $p_0$. The longest path in a
VCube with $n$ processes is $\log_2 n$. Therefore, when $n$ is small, the time to
send the message via the longest path is greater than the time
to send the $n-1$ messages sequentially using the ALL strategy. Considering
that the transmission time for each message is $t_s=0.1$, the interval between sending a
TREE message and receiving the corresponding ACK via the longest path
in the tree is $2\log_2 n (t_s+t_r+t_t)$. In the ALL strategy, to send $n-1$ messages, $(n-2)t_s+t_t+tr$ time units are required.

\pgfplotstableread[col sep=tab,header=true]{tests/test.all.dat}\tablatencyAll
\pgfplotstableread[col sep=tab,header=true]{tests/test.root.dat}\tablatencyRoot
\pgfplotstableread[col sep=tab,header=true]{tests/test.flood.dat}\tablatencyFlood

\begin{figure}[htb]
\centering
  \subfigure[Latency] {\label{fig:graph.ff.lat}%
\begin{tikzpicture}[scale=0.6]
\begin{semilogxaxis} [
   xlabel = Processes,
   xtick=data,
   ymajorgrids,
   xticklabels from table = {\tablatencyAll}{p},
   ylabel = Latency (time),
   legend style={font = \scriptsize,
                 cells={anchor=west},
                legend pos = north west},
]
\addplot[color=blue, mark=*] table[x = p, y = latency] from \tablatencyRoot ;
\addlegendentry{ATREE}
\addplot[color=red, mark=square*] table[color=red, mark=*, x = p, y = latency]
from \tablatencyAll ;
\addlegendentry{ALL}
\addplot[color=green, mark=triangle*] table[color=green, mark=o, x = p, y =
latency] from \tablatencyFlood ;
\addlegendentry{NATREE}
\end{semilogxaxis}
\end{tikzpicture}
}
  \hfil
  \subfigure[Throughput] {\label{fig:graph.ff.tput}%
\begin{tikzpicture}[scale=0.6]
\begin{semilogxaxis} [
   xlabel = Processes,
   xtick=data,
   ymajorgrids,
   xticklabels from table = {\tablatencyAll}{p},
   ymax = 0.5,
   ylabel = Throughput (msg/time unit),
   legend style={font = \scriptsize,
                 cells={anchor=west},
                legend pos = north east},
]
\addplot[color=blue, mark=*] table[x=p, y = throughput] from \tablatencyRoot ;
\addlegendentry{ATREE}
\addplot[color=red, mark=square*] table[x=p, y = throughput] from \tablatencyAll
;
\addlegendentry{ALL}
\addplot[color=green, mark=triangle*] table[x=p, y = throughput] from
\tablatencyFlood ;
\addlegendentry{NATREE}

\end{semilogxaxis}
\end{tikzpicture}
}
\caption{Best-effort broadcast: all processes are correct.}
\label{fig:graph.ff}
\end{figure}
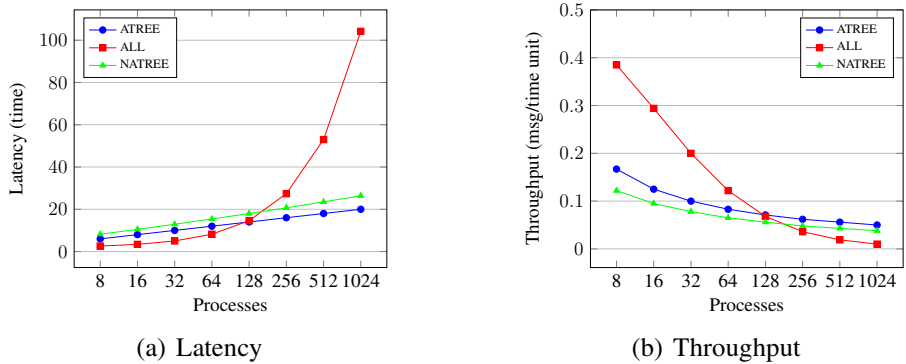

Thus, although ALL is more efficient for small systems, its performance declines as the system size grows. The NATREE non-autonomic tree solution 
presents similar behavior to ATREE, except for the extra latency that caused by processes having to wait for ACK/NACK messages during tree setup. In fault-free scenarios and after the tree is complete, the performance of both solutions is the same.

Throughput was computed as $1/latency$, since only one
process sends a single message. Similar to the latency, the ALL solution presents the best
performance for small systems, up to 128 processes. However, as $n$ increases, its performance decreases quickly. These results confirm the scalability of the proposed hierarchical strategy.

\begin{figure*}[htb]
\centering
\subfigure[Latency] {\label{fig:graph.floodfaults.lat}%
\begin{tikzpicture}[scale=0.65]
\begin{axis}[
   ybar,
   bar width=5pt,
   xtick = data,
   ymajorgrids,
   symbolic x coords = {0,1,2,3,4,5,6,7,8},
   xlabel = Crashed Processes,
   ylabel = Latency (time),
   ymin = 0,
    legend style={at={(1.01,1)}, draw=none, font = \scriptsize, anchor=north
west},
]
\addplot [draw=black, fill=blue, error bars/.cd,
            y dir=both,
            y explicit] table[x=f, y = latency, y error = desvpad2]
from \tabTestFloodTree ;
\addlegendentry{ATREE}
\addplot [draw=black, fill=red, error bars/.cd,
            y dir=both,
            y explicit] table[x=f, y = latency, y error = desvpad2] from
\tabTestFloodAll ;
\addlegendentry{ALL}
\addplot [draw=black, fill=green, error bars/.cd,
            y dir=both,
            y explicit] table[x=f, y = latency, y error = desvpad2] from
\tabTestFloodFlood ;
\addlegendentry{NATREE}

\end{axis}
\end{tikzpicture}
}
\hfil
\subfigure[Total messages]
{\label{fig:graph.floodfaults.msgs}%
\begin{tikzpicture}[scale=0.65]
\begin{axis}[
   ybar stacked,
   bar width=5pt,
   xtick = data,
   ymajorgrids,
   symbolic x coords = {0,1,2,3,4,5,6,7,8},
   xlabel = Fault Processes,
   ylabel = Messages,
      ymin = 0,
       legend style={at={(1.01,1)}, draw=none, font = \scriptsize, anchor=north
west},   
]
\addplot [xshift=-6pt, draw=black, fill=blue ] table[x=f, y = TREE] from
\tabTestFloodTree ;
\addlegendentry{ATREE (TREE)}
\addplot [xshift=-6pt, draw=black, pattern=crosshatch,
pattern color=blue,  error bars/.cd,
            y dir=both,
            y explicit] table[x=f, y = ACK, y error = desvpad1] from
\tabTestFloodTree ;
\addlegendentry{ATREE (ACK)}

\resetstackedplots{(0,0) (1,0) (2,0) (3,0) (4,0) (5,0) (6,0) (7,0) (8,0)}
\addplot [draw=black, fill=red ] table[x=f, y = TREE] from
\tabTestFloodAll ;
\addlegendentry{ALL (TREE)}
\addplot [draw=black, pattern=crosshatch,
pattern color=red,  error bars/.cd,
            y dir=both,
            y explicit] table[x=f, y = ACK, y error = desvpad1] from
\tabTestFloodAll ;
\addlegendentry{ALL (ACK)}

\resetstackedplots{(0,0) (1,0) (2,0) (3,0) (4,0) (5,0) (6,0) (7,0) (8,0)}

\addplot [xshift=6pt, draw=black, fill=green] 
	table[x=f, y = TREE] from \tabTestFloodFlood ;
  \addlegendentry{NATREE (TREE)}
\addplot [xshift=6pt, draw=black, pattern=crosshatch,
	pattern color=green] table[x=f, y = ACK] from \tabTestFloodFlood ;
  \addlegendentry{NATREE (ACK)}
\addplot [xshift=6pt, draw=black, pattern=crosshatch dots,
	pattern color=green,samples=500,
	    error bars/.cd,
            y dir=both,
            y explicit] 
        table[x=f, y = NACK, y error = desvpad1] from \tabTestFloodFlood ;
    \addlegendentry{NATREE (NACK)}
\end{axis}

\end{tikzpicture}}
\caption{Broadcast for $n=512$ varying the numbers of crashed processes.}
\end{figure*}
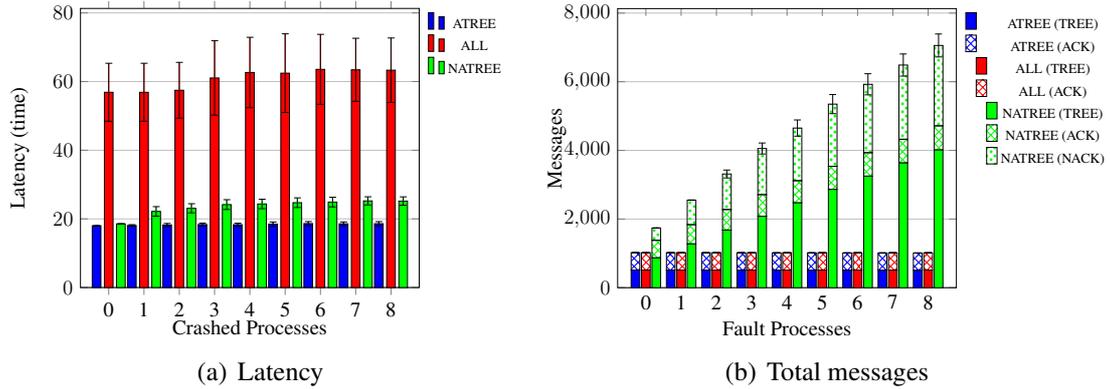

\noindent{\bf Experiments in scenarios with failures.} Scenarios with failures were executed in a system with 512 processes. The the number of crashed processes varied from 1 to $\log_2 n$. For each number of crashed processes, 100 different scenarios were generated randomly following a normal distribution. Unlike the scenarios without failures, in which a single message is sent, the experiments with failures consisted of broadcasting 10 messages. The average latency is shown in Figure~\ref{fig:graph.floodfaults.lat}. The ALL solution presents higher latency than ATREE because it needs to send all messages
sequentially, as explained above. In the NATREE solution, the result is similar to ALL since, for each failure, the tree needs to be rebuilt using flooding. ALL is more efficient both in the scenarios without and with few failures, but its performance degrades quickly as the number of failures increases. Regarding the number of messages, the results for ATREE and ALL are very similar, as shown in Figure~\ref{fig:graph.floodfaults.msgs}. In the case of NATREE, the total number of messages sent is higher, as it is necessary to rebuild the tree from the root using flooding. These results demonstrate the efficiency of the proposed solution in creating and maintaining the tree autonomically.

\pgfplotstableread[col sep=tab,header=true]
                  {tests/test.512.flood.tree.dat}\tabTestFloodTree
\pgfplotstableread[col sep=tab,header=true]
                  {tests/test.512.flood.flood.dat}\tabTestFloodFlood
\pgfplotstableread[col sep=tab,header=true]
                  {tests/test.512.flood.all.dat}\tabTestFloodAll

\section{Related Work}\label{sec:related}
This section reviews work related both in terms of the construction of spanning trees in distributed systems and the VCube virtual topology. It is divided into two main subsections: the first focuses on spanning tree algorithms, while the second surveys VCube and its applications in distributed computing.

\subsection{Spanning Tree Algorithms}

Two of the most classic algorithms for building a minimum spanning tree from a graph are the algorithm by Kruskal and Joseph \cite{kruskal:1956} and Prim's algorithm \cite{prim:1957}. Kruskal's algorithm initially creates a forest in which each vertex is a tree. At each step, the trees are connected through the lowest-weight edges. Edges that do not connect two trees are discarded, thus avoiding cycles. In the end, a single connected component is generated, which is the minimum spanning tree. Prim's algorithm uses a different approach, employing minimal cuts to select the lowest-weight edges for inclusion in the tree.

Many distributed algorithms for building spanning trees are based on the centralized Kruskal-Joseph and Prim algorithms. The first of these was defined by Gallager et al. \cite{gallager:1983}. The process is similar to that used by Kruskal. Initially, each process is a tree. At each level, a node is elected as leader, and a minimum-weight edge connecting it to a process in another tree is added. The process is repeated until a single connected component is formed. The algorithm proposed by Dalal \cite{dalal:1987} uses Prim's algorithm to connect tree segments by choosing the lowest-weight edge connecting two segments.

Avresky \cite{avresky:1999} presents three algorithms for constructing and
maintaining trees in fault-prone hypercube-based systems.
The algorithms tolerate single faults but may block in certain
combinations with multiple faults.

The work of Flocchini et al. \cite{flocchini:2012} proposes a fault-tolerant solution
for spanning trees that reconstructs the tree
after failures using pre-computed alternative trees considering all
failure possibilities.

\subsection{VCube Applications}
\label{sec:related_work_vcube}

VCube \cite{duarte2014vcube} was proposed as a scalable virtual topology for large-scale distributed systems, enabling efficient and fault-tolerant communication over a hierarchical structure. VCube ensures that both latency and the number of tests are provably bounded by logarithmic functions. VCube evolved from the earlier Hierarchical Adaptive Distributed System-level Diagnosis algorithm (Hi-ADSD)~\cite{duarte:1998}, that was applied for network fault management based on the SNMP (Simple Network Management Protocol)~\cite{duarte2002dependable}. Hi-ADSD presented a latency of $log_2^2n$ testing rounds. Another version called isochronous~\cite{brawerman2001isochronous} reduced the latency to the same of VCube $log_2 n$ rounds. VCube guarantees that latency by having each correct process test all $log_2 n$ clusters in all rounds. Hi-ADSD already exhibited several logarithmic properties, but in certain rare fault situations, it resulted in a quadratic number of failures. Later Hi-ADSD with Detours~\cite{duarte2009hierarquical} was proposed to guarantee a logarithmic number of tests in all situations, but the proof of the worst case remains open. DiVHA (Distributed Virtual Hypercube Algorithm) was the first hierarchical diagnosis algorithm to have a provably scalable number of tests~\cite{bona-overlays:2015}, and was implemented as an overlay network~\cite{bona2008hyperbone}, ensuring scalability and robustness under churn.

The original Hi-ADSD algorithm assumed static events, i.e. a process could only fail or recover after the previous event had been completely diagnosed. The algorithm presented in ~\cite{duarte2000algorithm} allows dynamic events, and furthermore allows a correct process to obtain diagnostic information from multiple other processes, employing timestamps to allow more recent events to be identified. 

Several broadcast algorithms have been designed based on the VCube. An autonomic reliable broadcast algorithm using dynamic spanning trees built within a VCube were presented in~\cite{rodrigues2014autonomic}. That algorithm was later extended to be able to work in an asynchronous systems in which it is impossible to determine whether a process has crashed or is simply slow~\cite{jeanneau2017autonomic}. Communication-efficient causal broadcast protocols based on  multiple intersecting trees built on a VCube were introduced in~\cite{araujo-brcast:2018} and later applied to implement a publish/subscribe system in VCube-PS~\cite{de2019vcube,de2017publish}. Message bundling techniques to reduce the communication overhead in VCube-based trees were explored in~\cite{rodrigues2018bundling}.

VCube has also been used as a substrate for autonomic spanning tree construction and self-adaptive distributed coordination. Distributed and autonomic minimum spanning tree algorithms over VCube were proposed in~\cite{rodrigues2014arvores} (\textit{in Portuguese}), which was later employed in several higher-level distributed algorithms such as distributed $k$-mutual exclusion~\cite{rodrigues2018distributed} and majority quorum systems~\cite{rodrigues-quorum:2016}.

More recently~\cite{duarte2022distributed,duarte2023missing}, VCube has be specified as an unreliable failure detector~\cite{chandra1996unreliable,reynal2005short,turchetti2015implementation,turchetti2016qos}. Those works provide a framework for unifying distributed diagnosis and unreliable failure detectors. An eventually perfect hierarchical failure detector (Diamond-P-VCube) was proposed in~\cite{Stein-diamond:2023}. A leader election algorithm based on the VCube assuming a crash-recovery model was addressed in~\cite{rodrigues-leader:2024}. Additionally, VCube have also been applied as a fault-tolerant parallel sorting method ~\cite{camargo:2024} for building fault-tolerant parallel algorithms.

Atomic broadcast has also been explored in the context of VCube. Leaderless and hierarchical atomic broadcast algorithms relying on autonomic VCube spanning trees were proposed in~\cite{Ruchel-leaderless:2023,Ruchel-jpdc:2024}. In addition, VCube has been adopted as a substrate for blockchain systems. The permissioned vCubeChain architecture leverages the VCube topology to provide scalable and fault-tolerant consensus mechanisms~\cite{freitas-vcubechain:2024}. Furthermore, efficient synchronization of conflict-free replicated data types (CRDTs) over VCube-based publish/subscribe systems has been proposed, demonstrating the suitability of the topology for large-scale eventual consistency services~\cite{Galesky-vcubeps:2023}.

\section{Conclusions}\label{sec:conclusions}

This work presented a solution for creating autonomic spanning trees in distributed systems subject to crash faults. The system's processes are organized in a logical topology called VCube. The trees are dynamically generated from a source process and reconfigure autonomically as processes fail. Two broadcast algorithms were implemented using the proposed tree strategy: the first for best-effort broadcast and the second for reliable broadcast. Simulation results are presented, demonstrating the efficiency of the proposed solutions and highlighting their scalability.

Future work includes extending the proposed algorithms to support the recovery of processes. Assuming a model with network partitions is also left as future work. Furthermore, an  implementation of a network service ~\cite{venancio2022nfv, venancio2019nfv,turchetti2017nfv} for message dissemination based on the proposed spanning tree algorithms that can be invoked by arbitrary applications over a network is also planned.

\section*{Acknowlegments}
This work was partially supported by the Brazilian Research Council (CNPq) Grant  305108/2025-5.

\bibliographystyle{IEEEtran} 
\bibliography{DistAutSpanTrees}

\end{document}